\magnification=1200
\def\qed{\unskip\kern 6pt\penalty 500\raise -2pt\hbox
{\vrule\vbox to 10pt{\hrule width 4pt\vfill\hrule}\vrule}}
\null\bigskip\bigskip
\centerline{ENTROPY PRODUCTION IN QUANTUM SPIN SYSTEMS.}
\bigskip
\centerline{by David Ruelle\footnote{*}{IHES.  91440 Bures sur Yvette,
France. $<$ruelle@ihes.fr$>$}.}
\bigskip\bigskip\noindent
	{\leftskip=2cm\rightskip=2cm\sl Abstract.  We consider a quantum spin system consisting of a finite subsystem connected to infinite reservoirs at different temperatures.  In this setup we define nonequilibrium steady states and prove that the rate of entropy production in such states is nonnegative.\par}
\vfill\eject
\null\bigskip\bigskip
	{\sl For several decades, Joel Lebowitz has been the soul of research in statistical mechanics.  He now plays a central role in the development of new ideas which reshape our understanding of nonequilibrium.  The present paper, dedicated to Joel on his 70-th birthday, extends some of the new ideas to quantum systems.}
\medskip
	{\bf Introduction.}
\medskip
	Consider a physical situation where a ``small'' system $S$ is connected to different ``large'' heat reservoirs $R_a$ ($a=1,2,\ldots$) at different inverse temperatures $\beta_a$.  We want to define nonequilibrium steady states for the total system $L=S+R_1+R_2+\ldots$, and verify that the rate of entropy production in such states is $\ge0$.  The model which we discuss in this paper is that of a fairly realistic quantum spin system.  In what follows we first describe the model and state our assumptions (A1), (A2), (A3).  In this setup we introduce nonequilibrium steady states $\rho$ as states which, in the distant past, described noninteracting reservoirs at different temperatures.  Under suitable conditions we check that our definition does not depend on where we place the boundary between the small system and the reservoirs.  Our definition of the entropy production $e_\rho$ also does not depend on where the boundary between the small system and the reservoirs is placed.  With this definition we prove $e_\rho\ge0$.  By contrast with an earlier paper [4], we omit here assumptions of asymptotic abelianness in time which are difficult to verify, the definition of nonequilibrium steady states is more general, but we obtain less specific results.  
\medskip
	{\bf Description of the model.}\footnote{*}{See [3], [1].}  
\medskip
	Let $L$ be a countably infinite set.  For each $x\in L$, let ${\cal H}_x$ be a finite dimensional complex Hilbert space, and write ${\cal H}_X=\otimes_{x\in X}{\cal H}_x$ if $X$ is a finite subset of $L$.  We let ${\cal A}_X$ be the C$^*$-algebra of bounded operators on ${\cal H}_X$, and if $Y\subset X$ we identify ${\cal A}_Y$ with a subalgebra of ${\cal A}_X$ by the map ${\cal A}_Y\mapsto{\cal A}_Y\otimes{\bf 1}_{{\cal H}_{X\backslash Y}}\subset{\cal A}_X$.  We write $L$ as a finite union $L=\cup_{a\ge0}R_a$, where $R_0=S$ is finite ({\it small system}) and the $R_a$ with $a>0$ are infinite ({\it reservoirs}).  We can then define the {\it quasilocal} C$^*$ algebras ${\cal A}_a$, ${\cal A}$ as the norm closures of 
$$	\bigcup_{X\subset R_a}{\cal A}_X\qquad,
	\qquad\bigcup_{X\subset L}{\cal A}_X	$$  
repectively.  Note that all these algebras have a common unit element ${\bf 1}$.  In this setup we assume that an {\it interaction} $\Phi:X\mapsto\Phi(X)$ is given such that $\Phi(X)$ is a selfadjoint element of ${\cal A}_X$ for every finite $X\subset L$.  Also, for each reservoir, we prescribe an inverse temperature $\beta_a>0$ and a state $\sigma_a$ on ${\cal A}_a$.  
\medskip
	{\bf The assumptions {\rm (A1), (A2), (A3)}.}
\medskip
	(A1) {\sl The interaction $\Phi$ satisfies
$$	||\Phi||_\lambda=\sum_{n\ge0}e^{n\lambda}
	\sup_{x\in L}\sum_{X\ni x:{\rm card}X=n+1}||\Phi(X)||<\infty      $$
for some $\lambda>0$.}
\medskip
	The importance of this assumption is that it allows us to equip ${\cal A}$ with a one-parameter group $(\alpha^t)$ of automorphisms\footnote{*}{See [1] Theorem 6.2.4 (or [3] Section 7.6).} defining a {\it time evolution}.  Introduce a linear operator $\delta:\cup_{X\subset L}{\cal A}_X\to{\cal A}$ such that 
$$	\delta A=i\sum_{Y:Y\cap X\ne\emptyset}[\Phi(Y),A]\qquad
	{\rm if}\qquad A\in{\cal A}_X      $$
If $A\in{\cal A}_X$, one checks that 
$$	||\delta^mA||
\le||A||e^{\lambda{\rm card}X}m!(2\lambda^{-1}||\Phi||_{\lambda})^m      $$
The strongly continuous one-parameter group $(\alpha^t)$ of $*$-automorphisms of ${\cal A}$ is given by
$$	\alpha^tA=\sum_{m=0}^\infty{t^m\over m!}\delta^mA      $$
if $A\in\cup_{X\subset L}{\cal A}_X$ and $|t|<\lambda/2||\Phi||_{\lambda}$.  (More generally one could take $A\in{\cal A}_\lambda$, where ${\cal A}_\lambda$ is defined in the Appendix).  Let 
$$	H_\Lambda=\sum_{X\subset\Lambda}\Phi(X)      $$
for finite $\Lambda\subset L$ and $A\in{\cal A}$.  Writing $\Lambda\to L$ if $\Lambda$ eventually contains each finite $X\subset L$, we have then 
$$	\lim_{\Lambda\to L}
	||e^{itH_\Lambda}Ae^{-itH_\Lambda}-\alpha^tA||=0      $$
uniformly for $t$ in compact intervals of ${\bf R}$.  
\medskip
	(A2) {\sl $\Phi(X)=0$ if $X\cap S=\emptyset$, $X\cap R_a\ne\emptyset$, $X\cap R_b\ne\emptyset$ for different $a,b>0$.}
\medskip
	Note that the description of the interaction $\Phi$ is somewhat ambiguous because anything ascribed to $\Phi(X)$ might also be ascribed to $\Phi(Y)$ for $Y\supset X$.  Condition (A2) means that in our accounting, if a part of the interaction connects two different reservoirs, it must also involve the small system $S$.
\medskip
	(A3) {\sl If $a>0$, let $\Phi_a$ be the restriction of the interaction $\Phi$ to subsets of $R_a$ and write 
$$	H_{a\Lambda}=\sum_{X\subset R_a\cap\Lambda}\Phi_a(X)
	=H_{R_a\cap\Lambda}      $$
Let also the interactions $\Psi_{(\Lambda)}$ be given such that 
$$	||\Psi_{(\Lambda)}||_\lambda\le K<\infty\eqno{(1)}      $$
and write
$$	B_{a\Lambda}=\sum_{X\subset R_a\cap\Lambda}\Psi_{(\Lambda)}(X)      $$
We assume that, for a suitable sequence $\Lambda\to L$,
$$	\lim_{\Lambda\to L}{{\rm Tr}_{{\cal H}_{R_a\cap\Lambda}}
	(e^{-\beta_a(H_{a\Lambda}+B_{a\Lambda})}A)\over
	{\rm Tr}_{{\cal H}_{R_a\cap\Lambda}}
	e^{-\beta_a(H_{a\Lambda}+B_{a\Lambda})}}=\sigma_a(A)      $$
if $A\in{\cal A}_a$: this defines a state $\sigma_a$ on ${\cal A}_a$, depending on the choice of $(\Psi_{(\Lambda)})$ and the sequence $\Lambda\to L$.  Furthermore we assume that for each finite $X$ there is $\Lambda_X$ such that $\Psi_{(\Lambda)}(Y)=0$ if $\Lambda\supset\Lambda_X$ and $Y\subset X$; therefore
$$	||[B_{a\Lambda},A]||=0\eqno{(2)}      $$
if $\Lambda\supset\Lambda_X$ and $A\in{\cal A}_X$.}
\medskip
	In particular we can take all $\Psi_{(\Lambda)}=0$.  Using (3) below, it is readily verified that $\sigma_a$ is a $\beta_a$-KMS state (see [2]) for the one-parameter group $(\breve\alpha_a^t)$ of automorphisms of ${\cal A}_a$ corresponding to the interaction $\Phi_a$.  [I do not know which of the $\beta_a$-KMS states can be obtained in this manner].
\medskip
	Note that the assumptions (A1), (A2), (A3) can be explicitly verified in specific cases.  From (A3) we obtain the following result.
\medskip
	{\bf Lemma.}
\medskip{\sl
$$	\lim_{\Lambda\to L}||e^{it(H_{a\Lambda}+B_{a\Lambda})}A
e^{-it(H_{a\Lambda}+B_{a\Lambda})}-\breve\alpha_a^tA||=0\eqno{(3)}      $$
for $a>0$, and
$$	\lim_{\Lambda\to L}||e^{it(H_\Lambda+\sum_{a>0}B_{a\Lambda})}A
e^{-it(H_\Lambda+\sum_{a>0}B_{a\Lambda})}-\alpha^tA||=0\eqno{(4)}      $$
uniformly for $t$ in compact intervals of ${\bf R}$.}
\medskip
	We prove (4).  Write $\alpha_\Lambda^tA=e^{it(H_\Lambda+\sum_{a>0}B_{a\Lambda})}Ae^{-it(H_\Lambda+\sum_{a>0}B_{a\Lambda})}$ and $\delta_\Lambda A=i[H_\Lambda+\sum_{a>0}B_{a\Lambda},A]$.  If $A\in\cup_X{\cal A}_X$ we see using (1) that
$$   \alpha_\Lambda^tA=\sum_{m=0}^\infty{t^m\over m!}\delta_\Lambda^mA   $$
converges uniformly in $\Lambda$ for $|t|<\lambda/2(||\Phi||_\lambda+K)$.  Using also (2), it is shown in the Appendix that $\delta_\Lambda^mA\to\delta^mA$ in ${\cal A}$ when $\Lambda\to L$.  Therefore 
$$	\lim_{\Lambda\to L}||\alpha_\Lambda^tA-\alpha^tA||=0      $$ 
when $A\in\cup_X{\cal A}_X$, uniformly for $|t|\le T<\lambda/2(||\Phi||_\lambda+K)$.  But the condition $A\in\cup_X{\cal A}_X$ is removed by density, and the condition $|t|\le T<\lambda/2(||\Phi||_\lambda+K)$ by use of the group property.  The proof of (3) is similar.\qed
\medskip
	{\bf The KMS state $\sigma$.}
\medskip
	The interaction $\sum_{a>0}\beta_a\Phi_a$, evaluated at $X$ is $\beta_a\Phi_a(X)$ if $X\subset R_a$ and 0 if $X$ is not contained in one of the $R_a$.  The corresponding one-parameter group $(\beta^t)$ of automorphisms of ${\cal A}$ has, according to (A3), the KMS state\footnote{*}{The state $\sigma$ corresponds to the inverse temperature $+1$ rather than the inverse temperature $-1$ favored in the mathematical literature.} $\sigma=\otimes_{a\ge0}\sigma_a$ where $\sigma_0$ is the normalized trace on ${\cal A}_0={\cal A}_S$.  In fact 
$$	\sigma(A)=\lim_{\Lambda\to L}
{{\rm Tr}_{{\cal H}_\Lambda}(\exp(-\sum_a\beta_a(H_{a\Lambda}+B_{a\Lambda}))A)
	\over{{\rm Tr}_{{\cal H}_\Lambda}
	\exp(-\sum_a\beta_a(H_{a\Lambda}+B_{a\Lambda}))}}\eqno{(5)}      $$
\medskip
	{\bf Nonequilibrium steady states.}
\medskip
	We call {\it nonequilibrium steady states} (NESS) associated with $\sigma$ the limits when $T\to\infty$ of 
$$	{1\over T}\int_0^Tdt\,(\alpha^t)^*\sigma      $$
using the $w^*$-topology on the dual ${\cal A }^*$ of ${\cal A }$.  With respect to this topology, the set $\Sigma$ of NESS is compact, nonempty, and the elements of $\Sigma$ are $(\alpha^t)^*$-invariant states on ${\cal A }$.
\medskip
	This definition generalizes that given in [4] where, under stringent asymptotic abelian\-nes conditions, the existence of a single NESS was obtained.
\medskip
	{\bf Dependence on the decomposition $L=S+R_1+R_2+\ldots$}\footnote{**}{This section and the following Proposition are in the nature of a technical digression, and may be omitted by the reader essentially interested in the positivity of the entropy production.}
\medskip
	Our definition of $\sigma$, and therefore of $\Sigma$ depends on the choice of a decomposition of $L$ into small system and reservoirs.  If $S$ is replaced by a finite set $S'\supset S$ and the $R_a$ by correspondingly smaller sets $R'_a\subset R_a$ one checks that (A1), (A2), (A3) remain valid.  If $\Phi'_a$ is the restriction of $\Phi$ to subsets of $R'_a$, the replacement of $\sum\beta_a\Phi_a$ by $\sum\beta_a\Phi'_a$ changes $(\beta^t)$ to a one-parameter group $(\beta'^t)$ and $\sigma$ to a state $\sigma'$.  These changes are in fact bounded perturbations covered by Theorem 5.4.4 and Corollary 5.4.5 of [1].  The map $\sigma\to\sigma'$ (of KMS states for $(\beta^t)$ to KMS states for $(\beta'^t)$) is nonlinear (as can be guessed from (5)) and therefore we cannot expect that ${1\over T}\int_0^Tdt\,(\alpha^t)^*\sigma'$ has the same limit as ${1\over T}\int_0^Tdt\,(\alpha^t)^*\sigma$ in general, but the deviation is not really bad.  The (central) decomposition of KMS states into extremal KMS states gives factor states.  If $\sigma$ is assumed to be a factor state, and $(\alpha^t)$ is asymptotically abelian, one finds that $\lim{1\over T}\int_0^Tdt\,(\alpha^t)^*\sigma$ does not depend on the decomposition $L=S+R_1+R_2+\ldots$, as the following result indicates.
\medskip
	{\bf Proposition.}
\medskip
	{\sl Using the above notation, assume that $\sigma$ is a factor state, and that
$$	\lim_{t\to\infty}||[\alpha^tA,B]||=0      $$
when $A,B\in{\cal A}$.  Then, when $T\to\infty$, 
$$	\lim{1\over T}\int_0^Tdt\,(\alpha^t)^*\sigma'=
	\lim{1\over T}\int_0^Tdt\,(\alpha^t)^*\sigma      $$}
\indent
	Let us introduce the GNS representation $({\cal H},\pi,\Omega)$ associated with $\sigma$ so that if 
$$	\rho=\lim{1\over T}\int_0^Tdt\,(\alpha^t)^*\sigma      $$
we have
$$	\rho(A)=\lim{1\over T}\int_0^Tdt\,(\Omega,\pi(\alpha^tA)\Omega)      $$
By restricting $T$ to a subsequence we may assume that in the weak operator topology
$$  \lim{1\over T}\int_0^Tdt\,\pi(\alpha^tA)=\bar A\in\pi({\cal A})''  $$
and by assumption we also have $\bar A\in\pi({\cal A})'$, hence $\bar A\in\pi({\cal A})'\cap\pi({\cal A})''=\{\lambda{\bf 1}\}$ since $\sigma$ is a factor state.
\medskip
	But we may write $\sigma'(\cdot)=(\Omega',\pi(\cdot)\Omega')$: this follows from the perturbation theory of [1] (see proof of Theorem 5.4.4).  We have thus
$$	\lim{1\over T}\int_0^Tdt\,\sigma'(\alpha^tA)
	=\lim{1\over T}\int_0^Tdt\,(\Omega',\pi(\alpha^tA)\Omega')      $$
$$	=\lim{1\over T}\int_0^Tdt\,(\Omega,\pi(\alpha^tA)\Omega)
	=\lim{1\over T}\int_0^Tdt\,\sigma(\alpha^tA)      $$
as announced.\qed
\medskip
	{\bf Entropy production.}
\medskip
	For finite $\Lambda\subset L$ we have defined 
$$	H_\Lambda=\sum_{X\subset\Lambda}\Phi(X)      $$
but $H_L$, $H_{R_a}$ do not make sense.  We can however define
$$	[H_L,H_{R_a}]=\lim_{\Lambda\to L}[H_\Lambda,H_{R_a\cap\Lambda}]
	=\lim_{\Lambda\to L}[H_\Lambda,H_{a\Lambda}]      $$
We have indeed
$$	[H_\Lambda,H_{a\Lambda}]
	=[H_\Lambda-H_{a\Lambda},H_{a\Lambda}]
	=[H_\Lambda-\sum_{b>0}H_{b\Lambda},H_{a\Lambda}]      $$
and (A2) gives
$$	H_\Lambda-\sum_{b>0}H_{b\Lambda}
	=\sum_{x\in S}\sum_{X:x\in X\subset\Lambda}
	{1\over{\rm card}(X\cap S)}\Phi(X)      $$
[implying the existence of the limit $\lim_{\Lambda\to L}(H_\Lambda-\sum_{b>0}H_{b\Lambda})=H_L-\sum_{b>0}H_{R_b}\in{\cal A}$].  Using (A1) we obtain
$$	||[\Phi(X),H_{a\Lambda}]||
\le2\lambda^{-1}||\Phi||_\lambda||\Phi(X)||e^{\lambda{\rm card}X}      $$
hence 
$$	\sum_{X\ni x}||[\Phi(X),H_{a\Lambda}]||
	\le2\lambda^{-1}||\Phi||_\lambda e^{\lambda}||\Phi||_\lambda      $$
and $[H_\Lambda,H_{a\Lambda}]$ has a limit $[H_L,H_{R_a}]\in{\cal A }$ when $\Lambda\to L$ with 
$$	||[H_L,H_{R_a}]||
	\le2{\rm card}S\lambda^{-1}e^\lambda||\Phi||_\lambda^2      $$
The operator 
$$	i[H_L,H_{R_a}]      $$
may be interpreted as the rate of increase of the energy of the reservoir $R_a$ or (since this energy is infinite) rather the rate of transfer of energy to $R_a$ from the rest of the system.  According to conventional wisdom we define the rate of entropy production in an $(\alpha^t)^*$-invariant state $\rho$ as 
$$	e_\rho=\sum_{a>0}\beta_a\rho(i[H_L,H_{R_a}])      $$
(this definition does not require that $\rho\in\Sigma$).
\medskip
	{\bf Remark.}
\medskip
	If we replace $S$ by a finite set $S'\supset S$ and the $R_a$ by the correspondingly smaller sets $R'_a\subset R_a$, we have noted earlier that (A1), (A2), (A3) remain satisfied.  As a consequence of (A1) we have 
$$	i[H_L,H_{R_a}-H_{R'_a}]
=\lim_{\Lambda\to L}i[H_\Lambda,H_{a\Lambda}-H'_{a\Lambda}]
=\lim_{\Lambda\to L}\delta(H_{a\Lambda}-H'_{a\Lambda})      $$
(where the operator $\delta$ has been defined just after (A3)), hence
$$	\rho(i[H_L,H_{R_a}-H_{R'_a}])=\lim_{\Lambda\to L}
	\rho(\delta(H_{a\Lambda}-H'_{a\Lambda}))=0      $$
{\it i.e.}, the rate of entropy production is unchanged when $S$ and the $R_a$ are replaced by $S'$ and the $R'_a$.  The reason why we do not have $\rho(i[H_L,H_{R_a}])=0$ is mathematically because $H_{R_a}$ is ``infinite'' ($H_{R_a}\notin{\cal A }$), and physically because our definition of $\rho(i[H_L,H_{R_a}])$ takes into account the flux of energy into $R_a$ from $S$, but not the flux at infinity.
\medskip
	{\bf Theorem.}
\medskip
	{\sl The entropy production in a {\rm NESS} is nonnegative, {\it i.e.}, $e_\rho\ge0$ if $\rho\in\Sigma$.} 
\medskip
	We have seen that 
$$	[H_L,H_{R_a}]=\lim_{\Lambda\to L}[H_\Lambda,H_{a\Lambda}]      $$
$$	=\lim_{\Lambda\to L}
	[H_\Lambda-\sum_{b>0}H_{b\Lambda},H_{a\Lambda}]      $$
Therefore, using (A3) and $[H_{b\Lambda}+B_{b\Lambda},\sum_{a>0}(H_{a\Lambda}+B_{a\Lambda})]=0$, we find
$$	\sum_{a>0}\beta_a[H_L,H_{R_a}]=\lim_{\Lambda\to L}
[H_\Lambda-\sum_{b>0}H_{b\Lambda},\sum_{a>0}\beta_aH_{a\Lambda}] $$
$$	=\lim_{\Lambda\to L}[H_\Lambda-\sum_{b>0}H_{b\Lambda},
	\sum_{a>0}\beta_a(H_{a\Lambda}+B_{a\Lambda})] $$
$$	=\lim_{\Lambda\to L}[H_\Lambda+\sum_{b>0}B_{b\Lambda},
	\sum_{a>0}\beta_a(H_{a\Lambda}+B_{a\Lambda})] $$
in the sense of norm convergence.
\medskip
	We also have, for some sequence of values of $T$ tending to infinity and all $A\in{\cal A}$,
$$	\rho(A)=\lim_{T\to\infty}{1\over T}\int_0^Tdt\,\sigma(\alpha^tA)
	=\lim_{T\to\infty}\lim_{\Lambda\to L}
	{1\over T}\int_0^Tdt\,\sigma(\alpha_\Lambda^tA)   $$
where, by (4),
$$	\alpha_\Lambda^tA=e^{it(H_\Lambda+\sum_{a>0}B_{a\Lambda})}
Ae^{-it(H_\Lambda+\sum_{a>0}B_{a\Lambda})}\to\alpha^tA\hbox{ in norm}      $$
when $\Lambda\to L$, uniformly for $t\in[0,T]$.
\medskip
	Write 
$$	H_{B\Lambda}=H_\Lambda+\sum_{a>0}B_{a\Lambda}      $$
$$	G_\Lambda=\sum_{a>0}\beta_a(H_{a\Lambda}+B_{a\Lambda})
	+\log{\rm Tr}_{{\cal H}_\Lambda}
	\exp(-\sum_{a>0}\beta_a(H_{a\Lambda}+B_{a\Lambda}))      $$
Then the entropy production is 
$$	e_\rho=\rho(i\sum_{a>0}\beta_a[H_L,H_{R_a}])
	=\lim_{T\to\infty}\lim_{\Lambda\to L}{i\over T}\int_0^Tdt\,
	\sigma(e^{itH_{B\Lambda}}
	[H_{B\Lambda},G_\Lambda]e^{-itH_{B\Lambda}})      $$
and the convergence when $\Lambda\to L$ of the operator $(e^{itH_{B\Lambda}}[H_{B\Lambda},G_\Lambda]e^{-itH_{B\Lambda}})$ is uniform for $t\in[0,T]$.  According to (A3) we may choose the $\Lambda$ tending to $L$ such that ${\rm Tr}_{{\cal H}_\Lambda}e^{-G_\Lambda}(\cdot)$ tends to $\sigma(\cdot)$ in the $w^*$-topology, hence
$$	e_\rho=\lim_{T\to\infty}\lim_{\Lambda\to L}{i\over T}\int_0^Tdt\,
	{\rm Tr}_{{\cal H}_\Lambda}(e^{-G_\Lambda}e^{itH_{B\Lambda}}
	[H_{B\Lambda},G_\Lambda]e^{-itH_{B\Lambda}})      $$
$$	=\lim_{T\to\infty}\lim_{\Lambda\to L}{1\over T}\int_0^Tdt\,
	{\rm Tr}_{{\cal H}_\Lambda}(e^{-G_\Lambda}{d\over dt}
	(e^{itH_{B\Lambda}}G_\Lambda e^{-itH_{B\Lambda}}))      $$
$$	=\lim_{T\to\infty}\lim_{\Lambda\to L}{1\over T}
	\big({\rm Tr}_{{\cal H}_\Lambda}(e^{-G_\Lambda}e^{iTH_{B\Lambda}}
	G_\Lambda e^{-iTH_{B\Lambda}})
	-{\rm Tr}_{{\cal H}_\Lambda}(e^{-G_\Lambda}G_\Lambda)\big)    $$
and the Theorem follows from the Lemma below, applied with $A=G_\Lambda$, $U=e^{iTH_{B\Lambda}}$ and $\phi(s)=-e^{-s}$.
\medskip
	{\bf Lemma.}\footnote{*}{As R. Seiler kindly pointed out to me, this lemma can be obtained readily from O. Klein's inequality
$$	{\rm tr}(f(B)-f(A)-(B-A)f'(A))\ge0      $$
where $A$, $B$ are hermitean and $f$ convex: take $B=UAU^{-1}$ and $\phi=f'$.}
\medskip
	{\sl Let $A$, $U$ be a hermitean and a unitary $n\times n$ matrix respectively, and $\phi:{\bf R}\to{\bf R}$ be an increasing function.  Then
$$	{\rm tr}(\phi(A)UAU^{-1})\le{\rm tr}(\phi(A)A)      $$}
\indent
	By adding a constant we may assume $\phi\ge0$.  We write the spectral decomposition of $A$ as 
$$	A=\sum_{j=1}^na_jE_j=a_1P_1+\sum_{j=2}^n(a_j-a_{j-1})P_j      $$
where $a_1\le\ldots\le a_n$ are the eigenvalue of $A$, $E_j$ the spectral projections, and $P_j=E_j+\ldots+E_n$.  Therefore 
$$	\phi(A)=\phi(a_1)P_1+\sum_{j=2}^n(\phi(a_j)-\phi(a_{j-1}))P_j      $$
and
$$	{\rm tr}(\phi(A)UAU^{-1})=\phi(a_1){\rm tr}(P_1UAU^{-1})
	+\sum_{j=2}^n(\phi(a_j)-\phi(a_{j-1})){\rm tr}(P_jUAU^{-1})      $$
We have
$$	{\rm tr}(P_jUAU^{-1})
	=\sum_{k=j}^n\sum_{l=1}^na_l{\rm tr}(E_kUE_lU^{-1})
	=\sum_{k=j}^n\sum_{l=1}^nc_{kl}a_l      $$
where $c_{kl}={\rm tr}(E_kUE_lU^{-1})\ge0$ and $\sum_{k=1}^nc_{kl}=1$, $\sum_{l=1}^nc_{kl}=1$.  Hence
$$	{\rm tr}(P_jUAU^{-1})=\sum_{l=1}^n(\sum_{k=j}^nc_{kl})a_l
	\le\sum_{l=j}^na_l      $$
because $0\le\sum_{k=j}^nc_{kl}\le1$, $\sum_{l=1}^n\sum_{k=j}^nc_{kl}=n-j+1$ and $a_1\le\ldots\le a_n$.  Finally
$$	{\rm tr}(\phi(A)UAU^{-1})\le\phi(a_1)\sum_{l=1}^na_l
	+\sum_{j=2}^n(\phi(a_j)-\phi(a_{j-1}))\sum_{l=j}^na_l
	=\sum_{j-1}^n\phi(a_j)a_j={\rm tr}(\phi(A)A)      $$
proving the Lemma.\qed
\medskip
	{\bf Remark.}
\medskip
	We have
$$	\sum_{a>0}\rho(i[H_L,H_{R_a}])=0      $$
because 
$$	-\sum_{a>0}\rho(i[H_L,H_{R_a}])
=\lim_{\Lambda\to L}\rho(i[H_\Lambda,H_\Lambda-\sum_{a>0}H_{a\Lambda}])
={d\over dt}\rho(\alpha^t\sum_{X:X\cap S\ne\emptyset}\Phi(X))|_{t=0}=0      $$
where we have used the fact that $\rho$ is $(\alpha^t)^*$-invariant.  In particular, in the case of two reservoirs
$$	0\le e_\rho=(\beta_1-\beta_2)\rho(i[H_L,H_{R_1}])      $$
so that if the temperature $\beta_1^{-1}$ is less than $\beta_2^{-1}$, {\it i.e.}, $\beta_1-\beta_2>0$, the flux of energy into $R_1$ is $\ge0$: heat flows from the hot reservoir to the cold reservoir.
\medskip
	{\bf Proving strict positivity of $e_\rho$.}
\medskip
	It is an obvious challenge to prove that $e_\rho\ne0$.  A natural situation to discuss would correspond to $R_a={\bf Z}^\nu$ and $\Phi_a$ translationally invariant.  But we need then $\nu\ge3$ as discussed in [4].  Indeed, for $\nu<3$ one expects a nonequilibrium steady state to be in fact an equilibrium state at a temperature intermediate between the original temperatures of the reservoirs.  Instead of a quantum spin system as described above, a gas of noninteracting fermions would probably be easier to treat first.
\medskip
	{\bf Appendix: the algebras ${\cal A}_\lambda$.}
\medskip
	The purpose of this Appendix is to complete the proof of (4) by establishing (10) below. 
\medskip
	If $\lambda\ge0$, a norm $||.||_\lambda$ is defined on $\cup_X{\cal A}_X$ by
$$	||A||_\lambda=\inf\sum_X||A_X||e^{\lambda{\rm card X}}      $$
where the $\inf$ is taken over finite decompositions $A=\sum_XA_X$ with $A_X\in{\cal A}_X$.  We let ${\cal A}_\lambda$ be the completion of $\cup_X{\cal A}_X$ with respect to that norm.  If $\lambda>\mu\ge0$ there are natural continuous maps ${\cal A}_{\lambda}\to{\cal A}_\mu\to{\cal A}$, which are not claimed to be injective.
\medskip
	If $||\Phi||_\lambda<\infty$ and $A_X\in{\cal A}_X$ the formula
$$	\delta A_X=i\sum_{Y:Y\cap X\ne\emptyset}[\Phi(Y),A_X]      $$
defines an element of ${\cal A}_\lambda$.  If $\lambda>\mu\ge0$, and $||\Phi||_\lambda<\infty$, one also checks that $\delta^m$ defines a map ${\cal A}_\lambda\to{\cal A}_\mu$ such that
$$	||\delta^mA||_\mu\le
||A||_\lambda m!(2(\lambda-\mu)^{-1}||\Phi||_\lambda)^m\eqno{(6)}      $$
\indent
	We have $\delta_\Lambda=\delta_\Lambda'+\delta_\Lambda''$, where
$$	\delta_\Lambda'A=i[H_\Lambda,A]\qquad,
	\qquad\delta_\Lambda''A=i[\sum_{a>0}B_{a\Lambda},A]      $$
and (1) and (6) (for $m=1$) yield
$$  ||\delta A||_\mu\le||A||_\lambda.2(\lambda-\mu)^{-1}||\Phi||_\lambda  $$
$$	||\delta_\Lambda'A||_\mu
	\le||A||_\lambda.2(\lambda-\mu)^{-1}||\Phi||_\lambda      $$
$$	||\delta_\Lambda''A||_\mu\le||A||_\lambda.2(\lambda-\mu)^{-1}K      $$
Given $\epsilon>0$ and $A\in{\cal A}_\lambda$ we can find $X$ such that $A=A_1+A_2$ with $A_1\in{\cal A}_X$ and $||A_2||_\lambda<\epsilon$.  Therefore 
$$	||(\delta-\delta_\Lambda)A||_\mu\le
	||(\delta-\delta_\Lambda)A_1||_\mu+||\delta A_2||_\mu
	+||\delta_\Lambda'A_2||_\mu+||\delta_\Lambda''A_2||_\mu      $$
$$	=||(\delta-\delta_\Lambda)A_1||_\mu
	+\epsilon.2(\lambda-\mu)^{-1}(2||\Phi||_\lambda+K)\eqno{(7)}      $$
Taking $\Lambda\supset\Lambda_X$ we also have  
$$	\delta_\Lambda''A_1=0      $$
by (2), and 
$$	(\delta-\delta'_\Lambda)A_1
	=i\sum_{Y:Y\not\subset\Lambda,Y\cap X\ne\emptyset}[\Phi(Y),A_1]      $$
so that
$$	||(\delta-\delta_\Lambda')A_1||_\mu
\le||A_1||_\lambda.2(\lambda-\mu)^{-1}||\Phi'||_{X\lambda}\eqno{(8)}      $$
where $||\Phi'||_{X\lambda}=\sup_{x\in X}\sum_{Y\ni x,Y\not\subset X}e^{({\rm card}Y-1)\lambda}||\Phi(Y)||$.  When $\Lambda\to L$ we have $||\Phi'||_{X\lambda}\to0$ and (7), (8) yield 
$$	\lim_{\Lambda\to L}||(\delta-\delta_\Lambda)A||_\mu=0\eqno{(9)}      $$
\indent
	We can now prove that, if $||\Phi||_\lambda<\infty$ and $A\in{\cal A}_\lambda$,
$$	\lim_{\Lambda\to L}||\delta^mA-\delta_\Lambda^mA||=0\eqno{(10)}      $$
We have indeed
$$	\delta^mA-\delta_\Lambda^mA
=\sum_{k=0}^{m-1}\delta_\Lambda^{m-k-1}(\delta-\delta_\Lambda)\delta^kA      $$
and, using (6), 
$$	||\delta^kA||_{2\lambda/3}
	\le||A||_\lambda.k!({6\over\lambda}||\Phi||_\lambda)^k      $$
hence, by (9),
$$  \lim_{\Lambda\to L}||(\delta-\delta_\Lambda)\delta^kA||_{\lambda/3}=0  $$
so that, using (6),
$$	||\delta_\Lambda^{m-k-1}(\delta-\delta_\Lambda)\delta^kA||
	\le||\delta_\Lambda^{m-k-1}(\delta-\delta_\Lambda)\delta^kA||_0      $$
$$	\le||(\delta-\delta_\Lambda)\delta^kA||_{\lambda/3}(m-k-1)!
	({6\over\lambda}||\Phi||_\lambda)^{m-k-1}      $$
which tends to zero when $\Lambda\to L$.  This concludes the proof of (10).
\medskip
	{\bf References.} 

[1] O. Bratteli and D.W. Robinson.  {\it Operator algebras and quantum statistical mechanics I, II.}  Springer, New York, 1979-1981 (2-nd ed. 1987-1997).

[2] R. Haag, N.M. Hugenholtz, and M. Winnink.  ``On the equilibrium states in quantum statistical mechanics.''  Commun. Math. Phys. {\bf 5},215-236(1967).

[3] D. Ruelle.  {\it Statistical mechanics.  Rigorous results.}  Benjamin, New York, 1969.

[4] D. Ruelle.  ``Natural nonequilibrium states in quantum statistical mechanics.''  J. Statist. Phys. {\bf 98},57-75(2000).

\end